# Scalar-Interchange Potential and Magnetic/Thermodynamic Properties of Graphene-like Materials

Elena Konstantinova, Ricardo Spagnuolo Martins, José Abdalla Helayel-Neto

Instituto Federal do Sudeste de Minas Gerais, Departamento de Educação e Ciências, 36080-001, Juiz de Fora, MG, Brazil
elena.konst@ifsudestemg.edu.br

Centro Brasileiro de Pesquisa Físicas, Rua Dr. Xavier Sigaud 150, 22290-180, Rio de Janeiro, RJ, Brazil
ricardosm@cbpf.br

Centro Brasileiro de Pesquisa Físicas, Rua Dr. Xavier Sigaud 150, 22290-180, Rio de Janeiro, RJ, Brazil
helayel@cbpf.br

## ABSTRACT

By means of numerical simulations, we explore possible influence of geometric deformations on magnetic graphene-like systems. As an efficient and economic method to take the geometrical effects into account, in addition to the electromagnetic interaction, we introduce an extra interaction transmitted by a massive scalar, associated to the Kekulé deformations. As a result, one meets a new potential interaction term, which affects the properties of the nano-surface. Monte Carlo analysis enables one to analyze the behavior of the system under variation of the external magnetic field, of the temperature, and also on the inverse of the mass of the extra scalar boson, which characterizes the typical length scale of geometric deformations. Our analysis is based on the spin configurations and includes evaluating magnetization, magnetic susceptibility and the specific heat in the presence of the Kekulé-induced new potential.

## 1 INTRODUCTION

Recently, graphene-like materials and their possible properties are raising a great deal of interest from the side of theoreticians, model-builders and experimentalists. In particular, the properties of graphene and graphene-like materials have been deeply inspected from both experimental and theoretical sides by means of different methods (see, e.g., [1, 2] for recent reviews). In general, it is important to have information about mechanical, electrical, magnetic and thermodynamic properties of different carbon surfaces. One of the interesting problems is related to the fact that the real graphene surfaces are not perfectly smooth and, therefore, their physical properties may depend on the geometry of the surface and on its deformations. Introducing curvature is a relevant step towards a clearer understanding of the connection between the possible surface geometries and the real physical properties of graphene-like structures. Nowadays, the effect of geometry in Condensed Matter Physics represents an active area of research and, in particular, one can observe an interest in inclusion of curvature effects to calculate physical properties of graphene-type materials [3, 4, 5, 6]. In the present paper, we propose to use one of the standard methods of Field Theory to describe the curvature of the surface. Since the two-dimensional gravitational interaction is known to be transmitted by a scalar field, we shall try to phenomenologically describe the effects of the curvature of the surface by means of a specially introduced massive scalar field. Then the inverse of the mass of such a field will characterize the typical length scale at which the effect of the curvature in a given point of the surface will propagate to achieve other points of the surface.

Let us note that the idea of introducing the geometric effects to the description of the two-dimensional surfaces is not really new. The distortions of the graphene planes may be attributed to the so-called Kekulé distortions [7, 8, 9], which are natural oscillations of the carbon bond lengths simultaneously stretching and compressing in alternating bonds. To account for a local Kekulé distortion different at each point of the graphene structure, we endow the distortions with the status of a scalar field to which we associate a mass-type parameter, as it shall become clear when our interaction Hamiltonian will be presented. We consider a spin-dependent effective interaction due to the exchange of the massive scalar field between the electrons. Our main purpose is to investigate how the new scalar- mediated interaction affects the physical properties of the system under consideration. As we shall see in brief, the particle-based description of the surface curvature leads to an additional interaction term in the Hamiltonian. Thus the calculations of physical properties in such a theory are not really difficult, such that the addition of the new term with a massive spin-boson exchange does not significantly increase the computational difficulties.

The paper is organized as follows. In Section 2 we describe our method of calculations, as well as the structure of the material and provide the criteria used for choosing the size of the system and the reference value for the choice of the mass of the particle which transmits the geometric interaction. Also, we carefully describe the Hamiltonian that governs the system under analysis and highlight the profile of the spin-dependent terms that stem from the scalar interchange between the electrons. In Section 3, we present the results of the calculations and discuss the thermodynamic properties of the system due to different possible values of the mass of the Kekulé fluctuations, such as the magnetization, specific heat and magnetic susceptibility. Finally, in Section 4, we cast our Concluding Comments.

## 2 METHOD OF CALCULATIONS

By using numerical methods, we intend to simulate the behavior of different classes of interaction potentials as a function of external and internal conditions of the system under investigation. Computational methods, in particular the Monte Carlo analysis, enable one to analyze the behavior of the systems in terms of variations of the external field, temperature, and the type of excitations that generate the potentials of self-interaction. We are actually contemplating situations that involve scalar, vector and tensor bosons, which may be representing a more fundamental physics behind the semi-microscopic approach [7-9]. In the present paper, we focus on the particular case of a massive scalar boson, referred to as the Kekulé scalar. It would be also interesting to assess the possibility that the gradient of the Kekulé scalar be associated to a massive vector field with a spin-dependent interaction that results from the interchange of the vectors. This shall be the subject of a further investigation. Here, however, we only contemplate the scalar case.

We assume that the graphene-like system is made out of a material with non-trivial magnetic properties. This means that each site of the graphene-like structure possesses some fixed magnetic moment, such that these magnetic moments interact between each other by the Heisenberg and dipole-dipole interactions. On the top of that, we assume that the curvature of the surface produce an extra contribution to the interaction, which can be described by a special new intermediate (scalar) boson, as specified below in Eq. (1).

For the numerical analysis of the magnetic structures described above, we have used the Monte Carlo simulations with the Metropolis algorithm [10, 11, 12]. The Metropolis Monte Carlo algorithm enables one to obtain the macro-state equilibrium for a physical system at the given temperature T. The basic idea of the method goes along the following procedure: we start off with some randomly chosen initial micro-state and then proceed by performing a very large number of random transformations of the micro-states, until we arrive at the equilibrium macro-state. In our case, we start the simulations with an initial configuration in which all spins have parallel directions. Then, the direction of one (randomly chosen) of these spins is randomly changed. In this way, we get to the new micro- and macro-states and evaluate the change of the overall energy compared to the previous configuration. If the energy variation is negative, $\Delta E < 0$, the temporary direction of the spin becomes permanent. If $\Delta E > 0$, the temporary direction becomes permanent with the probability $e^{-\Delta E/k_b T}$. We repeat this procedure a number of times equal to n = $10^4$ multiplied by a factor equal to the number of sites (spins). The final state corresponds to the stable configuration and is interpreted as the equilibrium macro-state. These preliminary calculations show that the equilibrium state is really achieved for $10^4$ Monte Carlo steps per spin and this number of steps is therefore adequate for our calculations. After that, all simulations have been performed for this choice of n.

By means of the method described above, we have explored the spin configurations, thermal equilibrium magnetization, the susceptibility and specific heat for the chosen structure. As a result of this study one can observe how these properties depend on the mass of the scalar boson. The Hamiltonian corresponding to the consideration presented above, can be cast in the form

$$H = -\vec{B} \cdot \sum_i \vec{S_i} - (J \sum_{<i,j>} \vec{S_i} \cdot \vec{S_j} + \tilde{J} \sum_{<i,j>} \frac{e^{-\xi r_{ij}}}{4\pi r_{ij}} \vec{S_i} \cdot \vec{S_j}) \quad (1)$$

$$-\omega \sum_{i<j} \frac{(\vec{S_i} \cdot \vec{e_{ij}})(\vec{e_{ij}} \cdot \vec{S_j}) - (\vec{S_i} \cdot \vec{S_j})}{r_{ij}^3} - \tilde{\omega} \sum_{i<j} \frac{(3 + 3\xi r_{ij} + \xi^2 r_{ij}^2)(\vec{S_i} \cdot \vec{S_j})(\vec{e_{ij}} \cdot \vec{S_j}) - (1 + \xi r_{ij})(\vec{S_i} \cdot \vec{S_j})}{4\pi \xi^2 r_{ij}^3}$$

In the first line, the double summations represents the ferromagnetic exchange between the nearest neighbors with a coupling constant, J. The first sum here stands for the coupling of the spins to an external magnetic field, B, and the last sum is the dipolar interaction term, where the coupling $\tilde{J}$ describes the strength of the dipole-dipole interaction. The second line contains a conventional dipole interaction term with strength $\omega$, and a dipole interaction term, $\tilde{\omega}$, due to the exchange of a massive scalar boson. The $S_i$'s are three-dimensional magnetic moments of unit length, $e_{ij}$ stands for the unit vectors pointing from the lattice site i to the lattice site j and $r_{ij}$ represent the distances between these lattice sites. The quantities $\omega$ and $\tilde{\omega}$ may be regarded as the coupling constants for the exchange term and the dipole-dipole interaction respectively. The parameter $\xi$ is the inverse mass of the scalar boson. We assume that the relation between the coupling constants is chosen such that $\omega/J = 0.001$, according to the paper [13].

Before starting to describe our simulations, let us first present some motivation for the introduction of the spin-dependent potential stemming from the scalar boson exchange. The whole idea is based on the work [14], where author studies interesting consequences of a local Kekulé distortion, that is, a Kekulé distortion that is different at each point of the plane. For this purpose, an extra scalar field is introduced. The role of the Kekulé scalar has been exploited in a great deal of details in the works of [15, 16, 17, 18].

Our point of view here is to propose that this scalar is a massive propagating degree of freedom that couples to the electrons and yields an effective interaction, which is spin-dependent and exhibits a screening parameter, $\xi$, that is nothing but the inverse mass of the exchanged Kekulé scalar. The explicit form of the spin-spin potential induced by the scalar exchange has been carefully worked out in the work by Dobrescu and Mocioiu [19]. So, our physical scenario relies upon the Kekulé scalar field as a way to take into account the non-smoothness of the graphene layers.

We assume that the external magnetic field is orthogonal to the plane of the structure. The simulations for the magnetization and magnetic susceptibility have been carried out for the values B = −20, −18,..., −8, −7.9, −7.8,..., 0,..., 7.8, 7.9, 8,..., 18, 20. Here, the energy and the applied magnetic field are expressed in units of J. The temperature is expressed in the units of J/kb, where J is the magnitude of the coupling constant and kb is Boltzmann's constant. In all these cases, the value of the temperature was chosen to be T = 0.2. In order to study the low-temperature thermodynamics, all simulations have been performed for temperatures essentially smaller than the critical temperature. The choice T = 0.2 provides a rapid convergence of the Monte Carlo procedure for the system of our interest.

We obtain the susceptibility, χ (in this case, along the OZ-axis), by using the Monte Carlo method, according to the expression

$$\chi = \frac{1}{k_b T N}(<m_z^2> - <m_z>^2) \qquad (2)$$

where N is the number of spins in the system and $<m_z>$ is the mean magnetization per spin in the z-direction. The specific heat, C, is obtained from the energy fluctuations relation

$$C = \frac{1}{k_b T^2 N}(<E^2> - <E>^2) \qquad (3)$$

where $<E>$ is the mean energy per spin. For calculating the specific heat we used B = 0 and the values T = 5.000, 4.975, 4.950,..., 0.050.

## 3 RESULTS AND DISCUSSIONS

Let us present the results of the Monte Carlo simulation for the structures of our interest and let us consider the thermodynamic behavior of the studied nanostructures. The thermal equilibrium results obtained by Monte Carlo simulations enable us to obtain the dependence for the specific heat and magnetic susceptibility versus the temperature. We notice that the calculations have been performed for values of the external magnetic field and temperature specified in Sec. 2. We have presented only part of the obtained data in the plots shown in Figs. 1-9, choosing such scales that the qualitative results become sufficiently explicit. The plots of the magnetic susceptibility in terms of the applied field for different values of ξ are presented in Figs. 1, 2 and 3

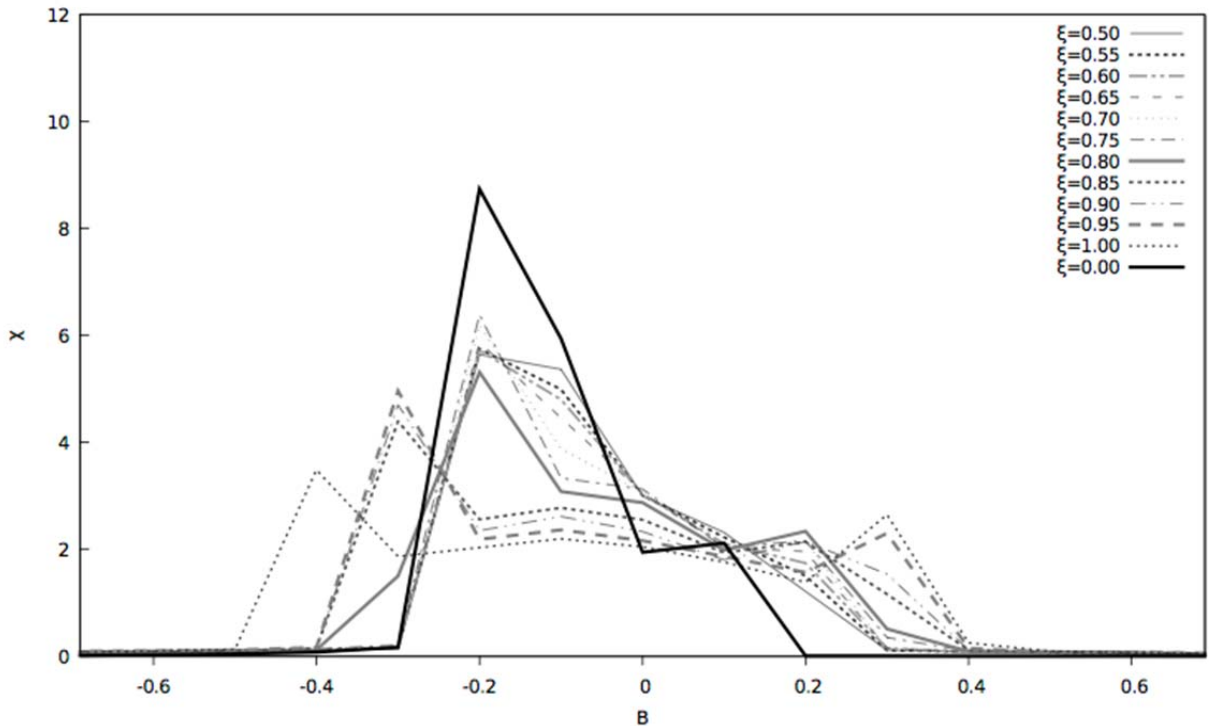

Figure 1: The plots of magnetic susceptibility versus applied field for all values of ξ.

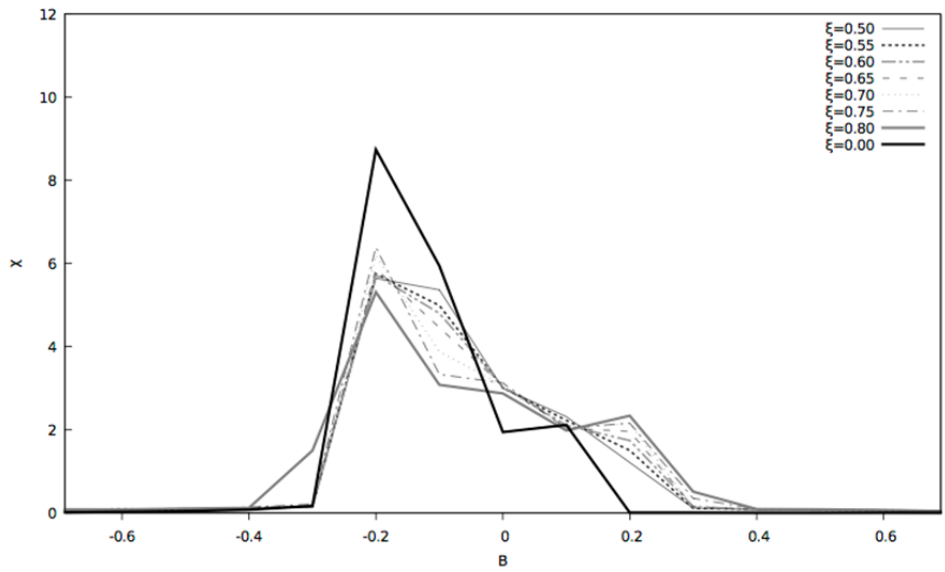

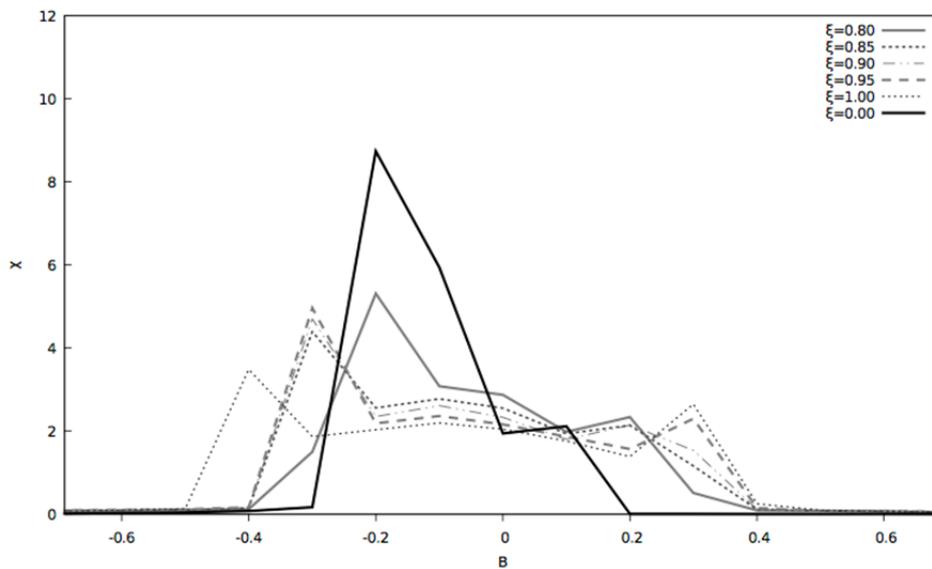

**Figure 2:** The plots of magnetic susceptibility versus applied field for values of ξ between 0.50 and 0.80.

**Figure 3:** The plots of magnetic susceptibility versus applied field for values of ξ between 0.80 and 1.00.

By analyzing the plots of Figs. 1, 2 and 3 one can conclude that the magnetic susceptibility of the studied nanostructure depends on the value of the mass of the scalar boson. The maximum of the curve falls on the same value of the external magnetic field for values of ξ from 0.50 to 0.75; for values from 0.80 to 1.00, the maximum of the curve shifts to the left, in the direction of increasing magnetic field. There are two peaks in the plots whose positions are shifted towards the reduced magnetic field values, with an increasing mass of the scalar boson between 0.80 to 1.00, and it remains constant for the smaller values of ξ.

If we keep in mind that ξ is the parameter which measures the screening of the potential, then the small values of ξ correspond to a very weak screening. This means a potential with a longer interaction range and, correspondingly, our plot discloses the result that for higher screenings the peaks shift more significantly whenever we change the applied external field.

Next, by analyzing Figs. 4, 5 and 6, one can understand how the critical temperature of the magnetic nanostructure depends on the value of ξ.

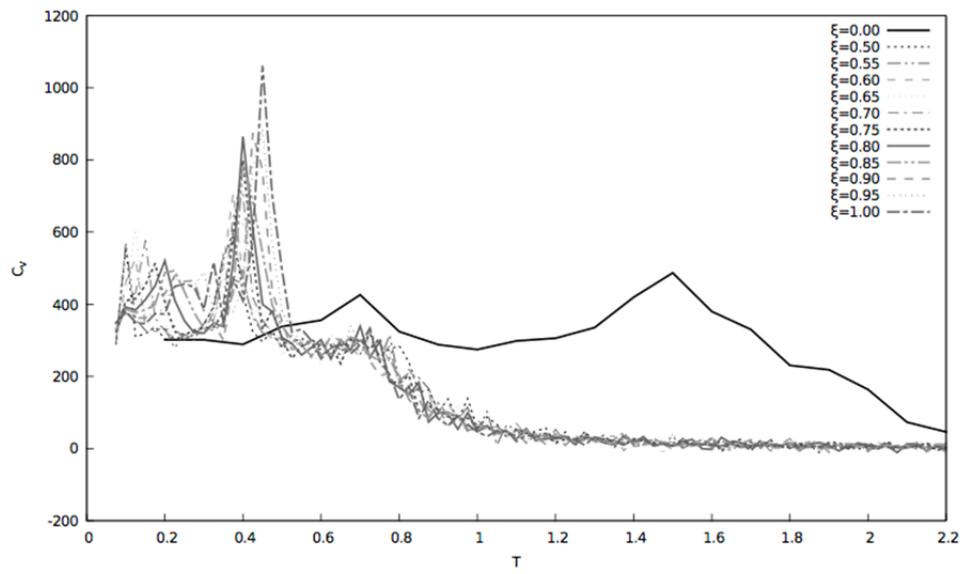

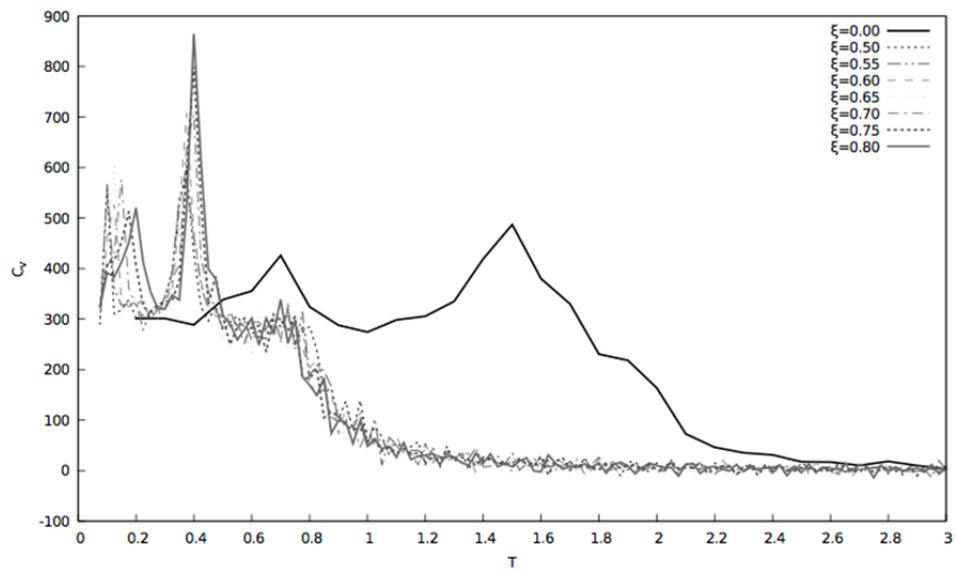

**Figure 4:** The plots of specific heat versus temperature without external magnetic field for all values of ξ.

**Figure 5:** The plots of specific heat versus temperature without external magnetic field for the values of ξ between 0.50 and 0.80.

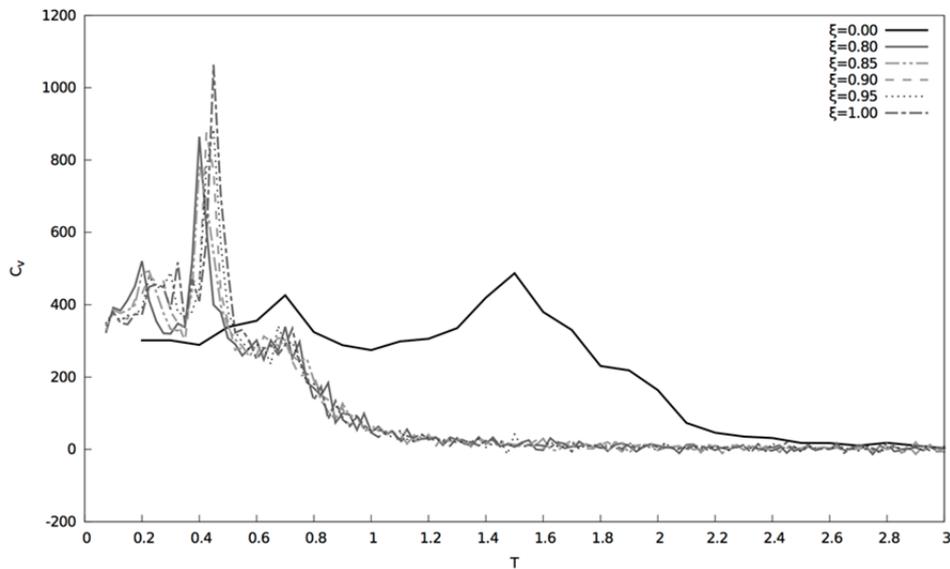

**Figure 6:** The plots of specific heat versus temperature without external magnetic field for the values of ξ between 0.80 and 1.00.

We conclude that the critical temperature grows with an increasing of the mass of the scalar boson. It can be seen that the line for the zero mass of the scalar boson, which corresponds to a perfectly smooth plane of graphene is very different from the lines where there are curvatures of the surface. The line corresponding to ξ = 0.0 on the plot is strongly shifted to the right relative to other lines, corresponding to the presence of the curvature of the surface. Moreover, one can observe the presence of two pronounced peaks, as shown in Fig. 4. It can be seen that with the increase of the mass of the scalar boson there is a shift in the direction of increasing temperature. Where in the first peak (with the temperature value close to 0.2), relative shift of the lines (to each other) is more than within the second peak, with the temperature value approximately 0.4. One can see that for the large values of the scalar boson mass, with the values of ξ varying from 0.85 to 1.00 (see Fig. 5), this shift is stronger than for the smaller values of mass, when ξ varies from 0.50 to 0.80. For that, see Fig. 6.

The plots of magnetization in terms of the applied magnetic field are shown in Figs. 7, 8 and 9. It can be noticed that the magnetization varies with increasing mass of the scalar boson in such a way that for larger values of ξ (corresponding to potentials with increasing screening effect), the magnetization changes more smoothly over a larger range of magnetic fields. The smaller is ξ, i.e. the larger is the range of the potential, then the magnetization plot tends to be more vertical and becomes less sensitive to the variation of the magnetic field.

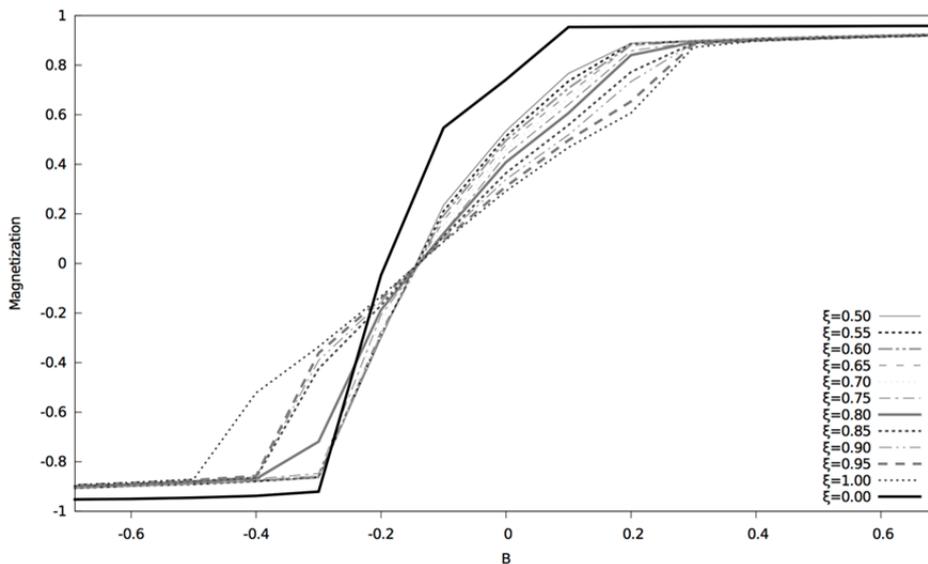

**Figure 7:** The plots of magnetization versus applied magnetic field for all values of ξ.

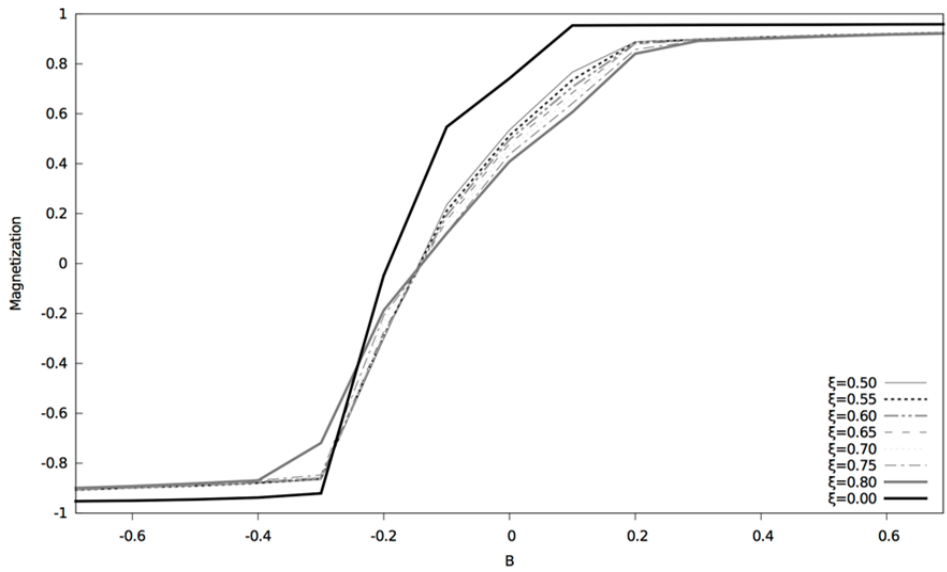

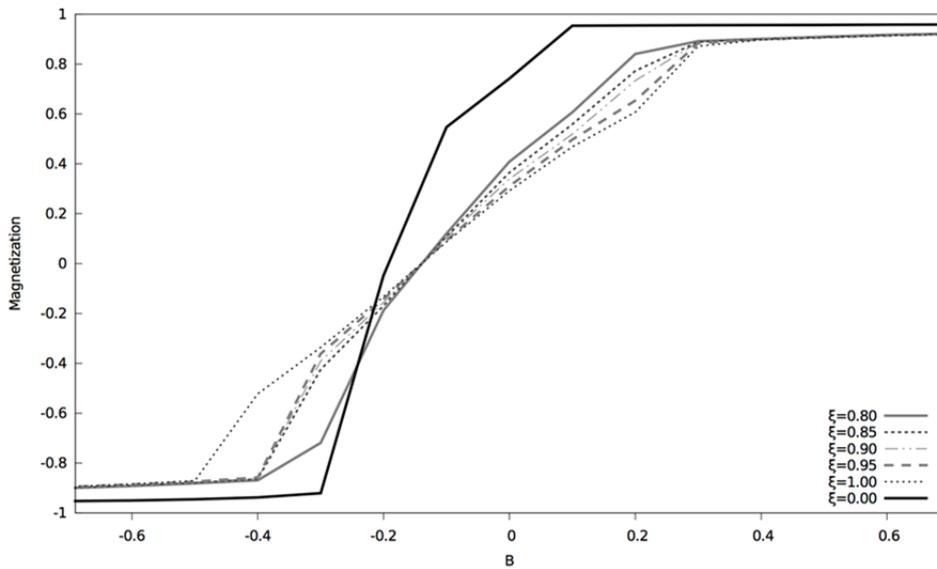

**Figure 8:** The plots of magnetization versus applied magnetic field for the values of ξ between 0.50 and 0.80.

**Figure 9:** The plots of magnetization versus applied magnetic field for the values of ξ between 0.80 and 1.00.

## 4 CONCLUSIONS

We have pursued an investigation of the magnetic and thermodynamic properties of a graphene-like system in a situation where the Kekulé deformations may be described by a massive scalar field. The interchange of this scalar gives rise to a screened spin-dependent potential and we were able to draw a number of conclusions about the parameter describing the mass-generated screening. We have pointed out how the specific heat, magnetic susceptibility and the magnetization behave under changes of ξ. It seems that the magnetic properties analyzed are more sensible on the range of the interaction (the mass of the Kekulé boson) than to the presence of the interaction with this foreign boson. Shorter ranges made these properties more sensible to the variations of the magnetic fields. In the magnetic susceptibility, greater values of the mass of the Kekulé boson changed the signature of the properties, significantly altering the difference in the intensity of both peaks observed in the graph and moving the peaks far apart from each other.

The specific heat property seems more sensitive to the interaction via Kekulé. When introduced the interaction, it immediately changes the signature of this property, moving the peaks closer together, increasing the difference in their intensities and showing the presence of a third unnoticed peak.

We have adopted that the electromagnetic interaction is present through its dipole-type contribution. At the same time, the Coulomb-type interaction now acquires the Yukawa-like form, since it is due to the massive scalar exchange. Finally, there is a spin-spin interaction, as it is given in the Hamiltonian of Eq. (1). Our general conclusion is that the Kekulé deformations may introduce sensitive changes in the physical properties of graphene, being even competitive with the purely electromagnetic interaction and contributing to increase the stability of the graphene structure against the electromagnetic repulsions. Our final conclusion is that the curvature effects of the surface can be described by means of a massive scalar field, providing an additional interaction term in the Hamiltonian. Using this method one can calculate the physical properties of the nanostructures. The practical use of the method is not difficult and does not significantly increase the computational time.

**AKNOWLEDGEMENTS**

Authors are grateful to CAT-CBPF for the computational facilities. R.S.M. and J.A.H.-N. express their gratitude to CNPq and FAPERJ for the financial support. E.K. is grateful to FAPEMIG for partial support through the "Universal" projects.